\documentclass[]{iopart}

\usepackage{epsfig}
\usepackage{setstack}

\def\nostrocostrutto#1\over#2{\mathrel{\mathop{\kern 0pt \rlap
  {\raise.2ex\hbox{$#1$}}}
  \lower.9ex\hbox{\kern-.190em $#2$}}}
\def\lsim{\nostrocostrutto < \over \sim}   

\begin{document}

\title[Multiparticle Production]%
{Theory of Multiparticle Production in the Soft Region}

\author{Valery A Khoze\dag\ and Wolfgang Ochs\ddag  
}

\address{\dag\ Department of Physics and Institute for
Particle Physics Phenomenology University of Durham,
Durham, DH1 3LE, UK }

\address{\ddag\  Max-Planck-Institut f\"ur Physik
(Werner-Heisenberg-Institut)\\ 
F\"ohringer Ring~6, D-80805 Munich, Germany}

\eads{\mailto{V.A.Khoze@durham.ac.uk},\mailto{ wwo@mppmu.mpg.de}}

\begin{abstract}
We review theoretical and experimental advances in the application
of the perturbative QCD approach to the description of particle
distributions
in jets in the (semi) soft region with particular emphasis on HERA physics.

\end{abstract}




\section{Introduction}

 In the last years hadron-jet physics has been intensively
studied in $e^+ e^-$, hadron-hadron and $ep$ scattering
processes.
%
The advantage of HERA is that here the measurements
can be performed at various hardness scales, from very large
$Q^2$ down to moderate and even small momentum transfer, thus
probing the interface between the hard and soft physics. 
The data clearly show that the broad features of hadronic jet
systems,
calculated at the parton level agree
surprisingly well with the measured ones.  This
proves the dominant role of the perturbative phase of jet
evolution and supports the hypothesis of local parton-hadron
duality (LPHD) \cite{dt,adkt1}. 
Important tests of the hadroproduction
dynamics come from the
studies of the so-called semisoft
region (small values of $x_p = p/E_{\rm jet}$,
but
formally $p \gg m_h$).
As long as the process
of colour blanching and hadronization 
is local in the configuration space  the
asymptotic shapes of the distributions are fully predicted.
The sub-asymptotic corrections are expressed
as power series in $\sqrt{\alpha_s}$. 
This is the so-called Modified Leading Logarithmic
Approximation
(MLLA) 
\cite{dt,ahm1} which takes into account
all essential ingredients of parton multiplication
in the next-to-leading order.

Colour coherence 
has led to a dramatic revision of
expectations for particle distributions. 
Thus, the coherent effects in the {\bf intrajet} 
cascades, resulting on the average, in the angular ordering
of sequential branching, gave rise to the {\it
hump-backed} shape of particle spectra
\cite{dfk1,bcmm}. It is not the
softest particles, but those with the  energies ($E_h
\sim E^{0.3-0.4}$) which multiply most effectively.
 Due to the {\bf interjet} coherence which is
responsible for the string \cite{ags}/drag \cite{adkt2} effects
in the multi-jet events, it can be verified,
that
the dynamics of the colour governs the hadroproduction.

In this talk we  focus on some selected
aspects of jet physics in
the (semi)soft region.  The main goal is to illustrate
some  phenomenological
advances of the perturbative approach,
for a broader outline see, e.g.\cite{dkmt1,ko}.


%

%
%

\section{Inclusive Spectra and Jet Universality}

A bright  
perturbative prediction
is the {\it hump-backed} shape of the partonic
distribution in the variable $\xi = {\rm log} \; \frac{1}{x}$. 
 As a 
consequence of coherence soft parton multiplication
is suppressed, and the spectrum 
acquires a characteristic bell-like shape \cite{dfk1,bcmm}.
It is amazing, that
the inclusive spectra of hadrons exhibit the same shape 
in a large kinematic region in a good agreement with the 
MLLA-LPHD predictions.  
Moreover, the data 
on various hard processes
demonstrate a remarkable universality assuming the
proper choice of the cascading  variables.
Recall that 
 the evolution parameter corresponding to the struck quark in DIS in
the Breit frame is  $\sqrt{Q^2}/2$ at four-momentum transfer $Q^2$.
The energy scale for  particle distribution in jets of energy $E_{jet}$ 
within the restricted cone $\Theta_0$ 
is $E_{jet}\Theta_0$ in the small angle approximation, see \cite{dkmt1}.
The $\xi$-distribution can also be discussed in terms of the moments or other
characteristics.
A summary of predictions and comparison with data can be found in
\cite{ko}. Here we discuss two recent results: the $\xi$-distributions at
HERA and the variation of the peak position  $\xi^*$
in the full energy range explored so far.

\subsection{$\xi$ distributions at HERA}

\begin{figure}[t]
\begin{center}
\mbox{\epsfig{file=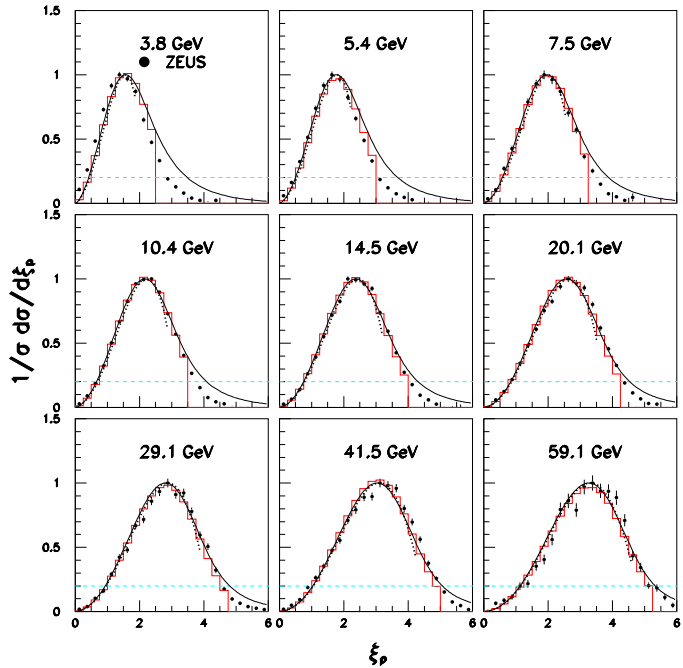,width=10cm}}
          \end{center}
\vspace{-0.3cm}
\caption{\label{zeusbrook}
The charged particle $\xi_p$ distribution in the current fragmentation
region in the Breit frame for different $Q$ bins as measured by the
ZEUS Collaboration \protect\cite{zeusxi}. 
The full and dotted lines
are the MLLA-M and MLLA-0 predictions with $\Lambda=280$ MeV respectively,
the histogram is the fit of the MLLA with $\Lambda=280$ MeV and $m_{eff}=230$
MeV, Figure from  
\protect\cite{bs}.}
\end{figure}

A recent measurement of the $\xi$ distribution by the ZEUS collaboration is
shown in \Fref{zeusbrook} \cite{zeusxi,bs}. The Gaussian shape is observed again at all
jet energies $E_{jet}=Q/2$. The MLLA prediction is based on leading and
next-to-leading order logarithmic approximations which become increasingly
better at higher jet energies and particle momenta exceeding the
cut-off $Q_0$ in the cascade evolution. 
A rather simple analytic expression is available in the
limit when the cut-off  and the QCD scale  become equal,
$Q_0=\Lambda$ (``limiting spectrum''). Because of the 
cut-off
$Q_0$ the spectrum ends at $\xi_{max}=Y=\log(E_{jet}/Q_0)$. The limiting
spectrum
(dotted line in \Fref{zeusbrook}, MLLA-0) 
ends at a finite value above the peak and fits the data rather well,
however, it predicts zero intensity 
beyond the limit $\xi_{max}$. 
In fact, the observed spectra in the variable $\xi_p=\log(E_{jet}/p)$ 
have no upper limit
for $p\to 0$, therefore in the very soft region with $p\sim Q_0$  
there is no agreement between theory and experiment.

This kinematical mismatch between theory and experiment can be avoided  
 if $Q_0$ is interpreted as particle mass and the theoretical prediction is
taken for the parton energy ($E=\sqrt{p^2+Q_0^2})$, or, 
$\xi_E=\log(E_{jet}/E)$ \cite{lo,klo}.
Then, both the theoretical prediction and the experimental data have the same
upper boundary $\xi_{max}=Y$. Transforming back the limiting spectrum to $\xi_p$ 
yields the prediction MLLA-M in \Fref{zeusbrook} which fits the full
spectrum quite well at the highest available energies but exceeds the data
at lower $Q$. An improved but still not fully satisfactory
result is obtained by treating the effective mass
as an additional parameter (histogram in \Fref{zeusbrook}). 
 
The discrepency of MLLA-M at large $\xi$ can be
attributed to the approximate form of the limiting spectrum 
obtained in MLLA which is not appropriate for the very soft particles
where the coupling $\alpha_s(k_T)$ becomes large. A better approximation
for the soft particles around $\xi\sim Y$ has been proposed
already some time ago \cite{lo} based on a modified   
perturbative expansion,
 and this analytic result has been shown to describe very well the $\xi$ 
spectrum above the
peak. These mass effects can clearly not be derived
directly from pQCD but they are plausible and
not in a contradiction either. Fortunately, with increasing jet energy these
kinematic effects become less and less important as already is visible in
\Fref{zeusbrook}.

Similar effects appear in the determination of the $\xi_p$ moments
which involve the integrals up to the upper boundary of the spectra. 
The results
at the lower energies $Q$ depend considerably on whether the cut-off
 $\xi_{max}$ 
is taken into account or not. This leads to large discrepencies between the
experimental $\xi_p$ moments \cite{zeusxi,bs} and the predictions from the
limiting spectrum \cite{dkt5} at low $Q$. On the other hand, 
the agreement is rather good if the 
moments are determined from the $\xi_E$ spectrum taking the particle mass 
as $Q_0$ \cite{lo}. Again, the results from different assumptions on 
mass effects converge with increasing $Q$ \cite{bs}.

Note that in spite of the known similarity
between the space- and time-like evolution at large $x$, there is an essential
difference between the small-$x$ behaviour of the DIS structure functions and the
jet fragmentation. In the former case at small Bjorken-$x$ 
the perturbative description
at very high energies
ultimately breaks down and 
the problem becomes essentially non-perturbative.
There are two main limitations on the application of the perturbative
approach: diffusion of gluons away from the hard scale into the infrared
and unitarity constraints,  see for example \cite{am4}. In the timelike
cascades, on the contrary, due to colour coherence 
the density of the low Feynman-$x$  particles
is not growing, and the influence of confinement dynamics reduces 
merely to an overall normalization.

Better salesmen might be tempted to claim that the observation
of the hump-backed plateau in jet fragmentation
has already clearly 
revealed the drastic low $x$-driven violation of DGLAP, a phenomenon which
in other observables 
many people are still so desperately targeting without any success.

\subsection{Peak position $\xi^*$ of $\xi$ distribution}
An important characteristic is
the peak position  $\xi^*$.
The high-energy behaviour of this quantity is predicted
\cite{dkt5} as
\begin{equation}
\xi^* \; = \; Y \left[ \frac{1}{2} + \sqrt{\frac{C}{Y}} - \frac{C}{Y} \right]
\label{eq:9}
\end{equation}
with the constant term given by
$$
C \; = \; \frac{a^2}{16 N_C b} \; = \; 0.2915 (0.3513) \;\qquad
 {\rm for} \; n_f \;
= \; 3(5),
$$
with $Y$ and $Q_0$ as given above.
Furthermore $ a = 11 N_C/3
+2 n_f/3 N_C^2$, $b = (11 N_C - 2 n_f) / 3$.
In the large-$N_C$-limit  parameter
$C$ becomes independent on both $n_f$ and $N_C$ and approaches its asymptotical
value of $C = \frac{11}{3} \frac{1}{2^4} \simeq 0.23$.  Therefore the energy behaviour
of $\xi^*$ is determined 
by such a fundamental
parameter of QCD as the celebrated $\frac{11}{3}$ factor 
in the coefficient $b$.
\begin{figure}
\begin{center}
\mbox{\epsfig{file=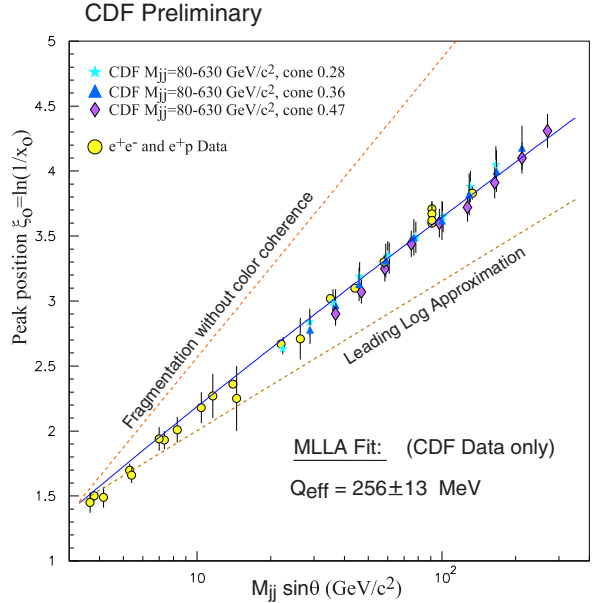,bbllx=0.5cm,bblly=4.2cm,bburx=19.0cm,%
bbury=23.4cm,width=8.cm}}  
          \end{center}
\vspace{-0.3cm}
\caption{\label{cdfxist}
Peak position $\xi^*$ of the inclusive $\xi$ distribution plotted against
di-jet mass $\times$ $\sin\Theta_0$ in comparison with the MLLA prediction
(central curve); also shown are (in arbitrary normalization) 
the double logarithmic approximation (lower
curve with asymptotic slope $\xi^*\sim Y/2$) and expectation from cascade
without coherence. Result by CDF Collaboration $\,$\protect\cite{cdfmult}.}
\end{figure}

A nice confirmation of the perturbative picture has been presented recently
by the CDF collaboration \cite{cdfmult}, see \Fref{cdfxist}.
 The studies of the charged particle 
 distributions were performed
for a variety of di-jet masses $(80 < M_{jj} < 630\ {\rm GeV}$)
and jet opening angles $\Theta_0$.  
The shapes of the measured $\xi$-distributions
at various hardness scales
$E_{jet} \Theta_0$ turn out to be very close to the
MLLA expectations.  As the di-jet mass increases, the peak position
$\xi^*$ shifts towards larger values of $\xi$ in a perfect agreement with the 
theory, whereas it joints smoothly towards 
the results from DIS and $e^+e^-$ collisions
at the lower energies.

 Note that by choosing a small opening angle one can study
relatively low hardness scales but in a more friendly environment:
because of the Lorentz boost, eventually all particles corresponding to the 
"hump" become relativistic. 
This is a serious argument against the attempts to explain the particle
depopulation at large $\xi$ by the finite mass effects.


\section{Soft Limit of Particle Spectra}

\begin{figure}[t]  
          \begin{center}
          \mbox{
\mbox{\epsfig{file=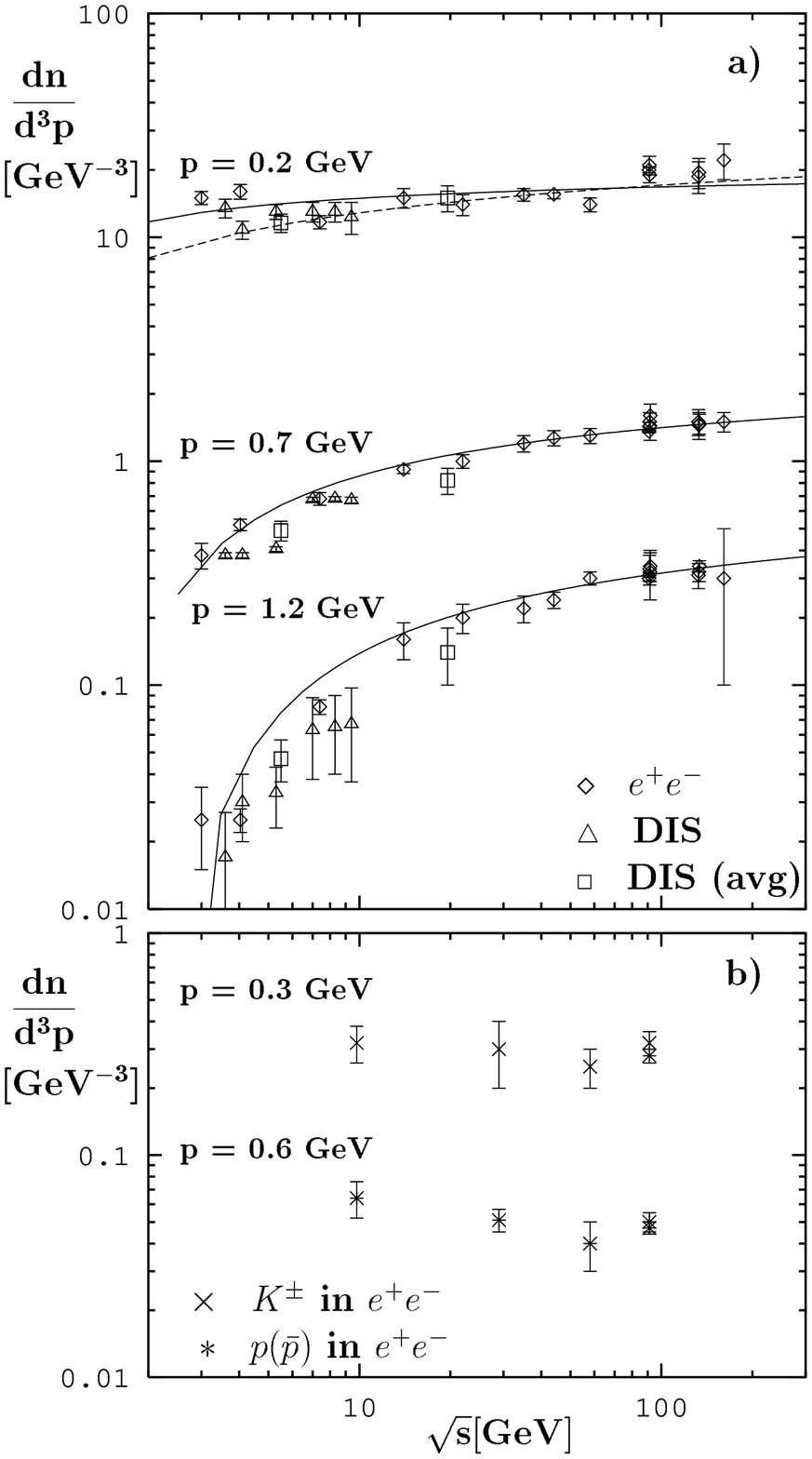,%
bbllx=2.5cm,bblly=12.6cm,bburx=16.5cm,bbury=28.0cm,%
height=8cm,clip=}}
}          \end{center}
\vspace{-0.2cm}
\hspace{5.5cm} {\bf $\sqrt{s}\ $ GeV}
\caption{\label{fig:dnd3p}
Particle density at fixed momentum $p$ as function of $cms$ energy
$\sqrt{s}$;
the dashed curve for an alternative implementation
of mass effects. Taken from~\protect\cite{klo1}.
}
\end{figure}

The low momentum data provide a
further confirmation of the basic ideas of QCD coherence and LPHD.
The analysis in \cite{klo1} was performed in terms of the 
particle density 
$ \frac{dn}{d^3 p}$ for $e^+ e^-$ annihilation  at low  momenta in
quite a wide $cms$ energy region (from ADONE to LEP-2).  The spectra were found
to be in a good agreement with analytical perturbative results applicable
for low particle energies emphasized above~\cite{lo,klo1}, see \Fref{fig:dnd3p}.
The  H1 data \cite{H1} also show agreement with the
perturbative expectations, thus confirming again the universality of soft particle
production.
 
The prediction for very low momenta $p\lsim$ mass is dependent to some
extent on the treatment of kinematic mass effects 
as discussed in the previous section. 
However, the fact that  in the soft limit $p\to 0$ 
the curves become flat in \Fref{fig:dnd3p}, 
i.e. independent of jet
energy, is a generic result and applies 
for different possible
implementations of mass effects studied in \cite{klo1}. 
It is a direct consequence of colour
coherence: in the soft limit the wave length of the produced particle
(gluon) is so large that it cannot resolve details of the primary parton jet
but probes only the primary $q\overline q$ pair. The emission probability
is then given by the Born term which is independent of $\sqrt{s}$.

We also emphasize the observation in \cite{klo1} that the approach to energy
independence for $p\to 0$ varies with particle type. It seems that the
energy independence is reached for momenta $p\lsim $ mass, namely for the
heavier protons already at much larger $p\sim 0.6$ GeV and for kaons at
$p\sim 0.3$ GeV.

Without doubt, it is not a priori clear at all, 
whether one can appeal to the perturbative expertise when exploring the low
momentum domain.  However, an attempt to stretch the perturbative expectations to
the limit of their applicability looks quite intriguing.  This could, in principle,
provide a clue for understanding of some conceptual problems of the LPHD.  
Whether or not the 
transition between two stages of jet development is smooth is a question
for experiment. It is therefore important to test these predictions 
further in various ways.

Another prediction of this kind for soft particles concerns the relative production in
quark and gluon jets. The same arguments -- dominance of Born term --
lead one to expect the ratio of densities 
$r(g/q)= \frac{dn}{d^3p}({g-jet})/\frac{dn}{d^3p}({q-jet})$ to approach the
limit~\cite{klo1} 
\begin{equation}
r(g/q) \to \ \frac{C_A}{C_F}=\frac{9}{4}\qquad
{\rm for}\quad p\to 0.
\label{eq:govq}
\end{equation}
This prediction would apply directly for the comparison
of soft particles in $q\overline q$ and $gg$ systems. The latter system can
be realized approximately in the process $e^+e^-\to q\overline q g$ with
the $q\overline q$ pair recoiling against the $g$; in this case the
ratio (\ref{eq:govq}) is found around $r(g/q)\sim1.8 $ \cite{opalg}, 
in a qualitative support of the
prediction (\ref{eq:govq}), but a bit lower than expected, and
this could be due to the non-collinearity of the three jets
in the actual measurement. Note that the ratio of total jet multiplicities
in quark and gluon jets stays considerably below the limit (\ref{eq:govq}),
both in theory and experiment, see for example review \cite{dg}.

Another interesting possibility to test the prediction 
(\ref{eq:govq}) is in the final
state of diffractive DIS to which we come back below.


\section{String/Drag Phenomena}

The existing data on  inter-jet particle flows in $e^+ e^- \rightarrow q\bar{q}g$ events
are also strongly in favour of soft gluon coherence.  
Within pQCD such 
effects arise from interference between the 
radiation off the $q, \bar{q}$ and $g$.
Note that colour-related  effects could play a valuable role
as a new tool helping to distinguish the new physics signals
\cite{dkmt1,ko}.

\subsection{Particle production between jets}
Let us recall a few facts concerning comparison with the theory
for the string/drag effect in $e^+ e^- \rightarrow q\bar{q}g$.
DELPHI~\cite{delphi4}
performed the  verification for the ratio $R_\gamma$
\begin{equation}
R_\gamma \; = \; \frac{N_{q\bar{q}} (q\bar{q}g)}{N_{q\bar{q}} (q\bar{q}\gamma)}
\label{eq:a77}
\end{equation}
of the particle densities in the interquark valley in the $e^+ e^- \rightarrow q\bar{q}g$ and 
$e^+ e^- \rightarrow q\bar{q}\gamma$ events 
for the $Y$-shaped  events using the $q$-jet tagging.  The 
ratio $R_\gamma$  in the $q\bar{q}$ angular 
interval [$35^\circ, 115^\circ$] was 
found to be
\begin{equation}
R_\gamma^{\exp} \; = \; 0.58 \: \pm \: 0.06.
\label{eq:a78}
\end{equation}
This is in a perfect agreement with the theoretical
expectation 
\cite{adkt2}.
\begin{equation}
R_\gamma^{th} \; \approx \; \frac{0.65 N_C^2 - 1}{N_C^2 - 1} \; \approx \; 0.61
\label{eq:a79}
\end{equation}

Another result \cite{delphi4} concerns the ratio $R_g = 
N_{qg}/N_{q\bar{q}}$ in the $q - g$ and $q - \bar{q}$ valleys
in the  symmetric $e^+ e^- \rightarrow 
q\bar{q}g$ events with the $q$-jet tagging.    
Quantitatively, 
comparing the minima  at $\pm [50^\circ, 70^\circ]$, this  ratio 
is found to be
\begin{equation}
R_g^{\exp} \; = \; 2.23 \: \pm \: 0.37.
\label{eq:a80}
\end{equation}
This is to be compared with the prediction 
\cite{adkt2}
\begin{equation}
R_g^{th} \; \approx \; 2.4.
\label{eq:a81}
\end{equation}

A clear  interference pattern 
arises in the  large-$E_T$ production of colourless objects, for instance, in 
$V +$ jet events
in proton scattering 
(with $V = \gamma, W^\pm$ or $Z$).  The hadronic antenna patterns for such processes are
analogous to that in the string-drag case.
Recently the first data on $W +$ jet production at the Tevatron
\cite{d00} have become 
available.  The interference effects are clearly seen, and they are in a good
quantitative agreement with the perturbative calculations \cite{ks}. 

Let us emphasize that in all drag-related measurements 
the inter-jet  flows are dominated by the low energy hadrons
(pions with typical momenta in the few 100 MeV range). It 
looks intriguing that such distant offsprings are controlled
by the pQCD rules.

\subsection{Particle production perpendicular to the event plane}

\begin{figure}[t]
\begin{center}
\mbox{\epsfig{file=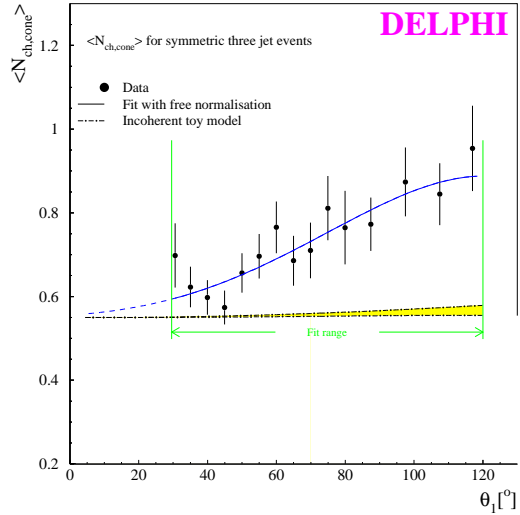,width=7.0cm
}}
          \end{center}
\caption{Multiplicity within a $30^\circ$-cone perpendicular to
the event plane of symmetric three-jet events
as a function of the angle $\Theta_1$ between the low energy jets
\protect\cite{delphi03}. The curve represents the
perturbative prediction (\ref{rperp}).}
\label{fig:del3jet}
\end{figure}

 An instructive test of the LPHD approach can be 
performed when studying the particle flow in the direction transverse 
to the 3-jet event plane,
see \cite{klo1}.  
Recently DELPHI~\cite{delphi03} have presented the first results on 
the particle yield 
in the transverse direction for the $Y$-shaped symmetric events.  
\Fref{fig:del3jet}
shows the multiplicity within the cone 
with the fixed half opening angle of $30^\circ$ perpendicular to the 
event plane. 
The plotted curve represents the lowest order formula
\begin{equation}
R_\perp=\frac{N_C}{4C_F}[2-\cos\Theta_{1+}-\cos\Theta_{1-} -\frac{1}{N_C^2}
(1-\cos \Theta_{+-})]  
\label{rperp}
\end{equation}
where the angles $\Theta_{ij}$ are between $q,\overline q, g$ 
(labeled as $+,-,1$).
We observe good
agreement with the angular dependence of the
perturbative prediction which provides a new test involving dominantly soft
particles, independent of the hadronization models.

The absolute normalization is also
possible in terms of the radiation into a cone of the same size in
two-jet events and this normalization is assumed in (\ref{rperp}).
It would  correspond to three jets with $\Theta_1=0$
in \Fref{fig:del3jet}. Changing the event
topology from the parallel $q - (qg)$
to anti-parallel $g - (q\overline q)$ configurations should increase the soft
perpendicular radiation by the factor 9/4
as in (\ref{eq:govq}) and (\ref{rperp}) for $\Theta_{+-}=0$, 
$\Theta_{1+}=\Theta_{1-}=\pi$. In \Fref{fig:del3jet} we are
observing just the beginning of this rise as far as is possible in 
this class of symmetric events. 

It would be interesting to measure the same radiation for other 
three-jet topologies closer to the $q\overline q - g$ configuration such as 
studied by OPAL \cite{opalg}. This would test equation (\ref{rperp})
closer to its maximum $R_\perp=N_C/C_F$. 

Furthermore, it would be interesting to see whether the increase of radiation
observed in \Fref{fig:del3jet}  holds down in the same way to very
low particle momenta. It seems impossible to have a directional dependence 
of densities in the
limit $p\to 0$, but it could become visible for momenta $p\sim$ mass. This is
the scale where the energy ($E_{jet}$) independence expected from the
perturbative calculation establishes for $\pi,K$ and $p$ particles \cite{klo1}.

\subsection{Conventional QCD vs. new physics mechanisms}

The antenna pattern can be 
used as a discriminative  tool to dissect the colour structure of the large-$E_T$ events, 
as a way to distinguish between the conventional QCD and  new physics mechanisms.
This diagnostic power is illustrated in \cite{hks} where 
the topology of hadronic 
flows corresponding to Higgs production at hadron colliders is discussed.
There the radiation samples were compared 
for the signal $(gg \rightarrow H \rightarrow 
b\bar{b} + g)$ and background $(gg \rightarrow b\bar{b} + g)$ production. 
It was found that the main difference between these two
came from the  radiation {\it between} the final-state jets. In particular,
there is, approximately 4/3 more radiation between these  jets for the Higgs 
production.  This is due to the absence of a colour  connection between the quarks in 
the background process.

Another topical example concerns 
central Higgs production in the events with double rapidity gaps,
for recent discussion and  references, see 
\cite{vk}.

\section{Further tests with soft particles in photoproduction 
and DIS}

Colour coherence leads to a rich 
diversity of effects 
in $\gamma p$ and DIS processes
(as well as in high-$p_\perp$  events 
in hadronic collisions). Here we note a few lines of further studies.

\subsection{Radiation perpendicular to event plane}

The dependence of the soft perpendicular radiation on the 
scattering angle of the hard partons in various di-jet production 
processes have been 
worked out \cite {klo}. 
In particular, for small angle scattering the intensity of soft particles
is larger in di-jet production with resolved photons (dominantly gluon
exchange) by the factor $C_A/C_F$ in comparison with direct photoproduction
(dominant quark exchange). 
Monte Carlo studies have shown that the expected
effects should be observable under realistic conditions \cite{bko}. 
The verification and detailed study
of the angular and momentum dependence
will provide further clues to the applicability of perturbative QCD
calculations in this regime.
%

\subsection{Soft radiation in diffractive 2-jet events} 
\begin{figure}[t]
\begin{center}
\mbox{\epsfig{file=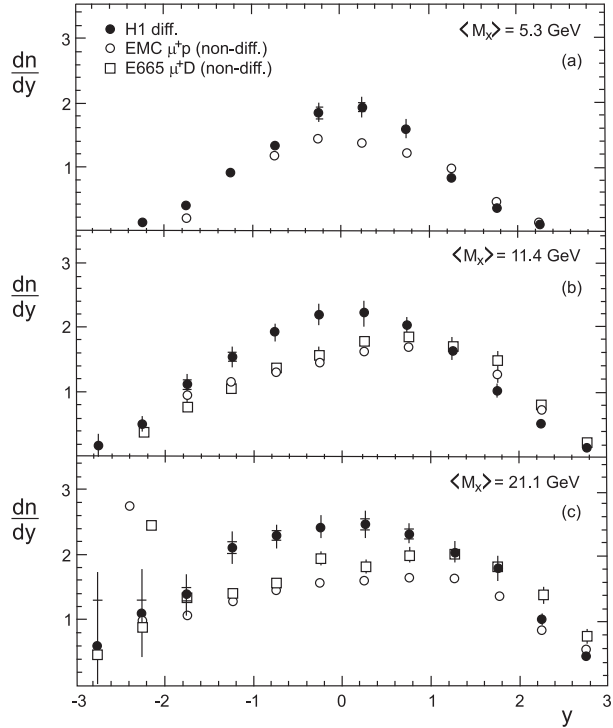,width=8.0cm%
}}
\end{center}
\caption{\label{fig:h1diff}
Rapidity distribution of charged particles in three intervals $M_X$ 
of the diffractively produced system in DIS (H1  Collaboration
\protect\cite{H1ydiff}) in comparison with results from
non-diffractive events at corresponding hadronic energies $W=M_X$. 
}
\end{figure}

In theoretical models the 
diffractive cluster produced in DIS is 
generated through Pomeron-photon interactions
leading to $q\overline q $ final state at low diffractive masses $M_X$
and to $ q\overline q g$ systems at higher masses
with the gluon in the Pomeron direction and
the $q\overline q $ pair in the photon direction. A recent jet analysis
by the ZEUS Collaboration \cite{zeusdj}, indeed, supports this view by showing 
the jet in the pomeron direction being broader than the one in photon
direction. Then the soft particles in this system are expected again to 
be produced by  colour octet sources and therefore the soft 
particle density is
higher. This phenomenon has been observed indeed by the H1 Collaboration
some time ago \cite{H1ydiff} in the comparison of rapidity distributions of
particles in the diffractive  and the non-diffractive systems, see
\Fref{fig:h1diff}. One observes the increased central production of
particles in the diffractive events as expected from the gluonic emitter. 
It is remarkable that the difference between the primary gluon and quark
jets appears already at such a low mass as $M_X\sim 5$ GeV. The difference
may become clearer if one restricts further to 2-jet events and  to
soft particles.

\subsection{Radiation in diffractive multi-jet events}
The multi-jet structure in diffractive DIS events opens up new possibilities
for tests of coherent soft particle production. In particular, the study of
diffractive three-jet events as in \cite{zeusdj} allows studies
complementary to the $e^+e^-$ case. For a given resolution $y_{cut}$
and sufficiently high diffractive mass $M_X$ 
the two-jet configuration should be
dominated by the octet configuration like $gg$, whereas the 3-jet
configuration contains the events with a gluon jet in Pomeron direction 
recoiling against $q\overline q$. The radiation into a cone perpendicular to
the 3-jet plane is then governed again by the general formula (\ref{rperp})
with the appropriate identification of angles. If normalized to 2-jet events
one would replace in (\ref{rperp}) the prefactor $N_C/C_F$ by 1 
corresponding to the replacement of the two-jet configuration $q\overline q$
by $gg$. 

It should be noted that particle density in the sidewise cone is not
entirely energy ($\sqrt{s}=M_X$) independent, this occurs only in the soft
limit when the lowest order radiation diagrams dominate.

The verification of the radiation pattern (\ref{rperp}) in diffractive
events would test both, the partonic interpretation of these
events, and the perturbative approach to the soft radiation.

\section{Summary}
The existing data  show that the analytical
perturbative approach to multihadron production in QCD jets is in a remarkably
healthy shape. This justifies the attempts to work in terms of quarks
and gluons down to the low momentum scale, the region which is usually
viewed as being entirely non-perturbative.
Further tests, especially of the 
characteristic predictions from soft
gluon coherence, should determine the range of applicability of the theory
towards  low momenta. 

It is clear that the agreement between parton level predictions and the data
cannot work in all aspects. Resonances
exist in nature but not in the quark gluon cascade. 
Recent studies by ZEUS \cite{zeuscor} show the qualitative agreement 
of perturbative predictions with data even for various 
multiparticle correlation phenomena  
in angular regions of variable size, but there is also a disagreement for correlations
between particles at low $p_T$ near the boundary $Q_0$.
In this kinematic region we encountered already some problems
in the inclusive studies of section 2. It is therefore important to explore
further the region of validity of this phenomenological framework.

The key concept of this approach is that the conversion of partons into hadrons
occurs at low virtualities, and that it is physics of QCD branchings 
which governs the
gross features of the jet formation.
The data demonstrate
that the transition betwen the perturbative and non-perturbative phases 
is quite smooth, and that the spectacular coherence phenomena successfully survive
the hadronization stage.

 Concluding this talk let us emphasize that, certainly, there is no mystery within
 the pQCD framework.
 Of real importance is that the experiment proves that the transformer
 between the perturbative and non-perturbative phases acts very smootly.
 This may (one day) shed light on the dynamics of colour confinement.
\ack
VAK thanks Leverhulme Trust for a Fellowship and the theory group of the
Max-Planck-Institute, Munich for their hospitality.

\end{document}